\documentclass[10pt, conference, compsocconf]{IEEEtran}

\ifCLASSINFOpdf

\else

\fi

\usepackage{cite}
\usepackage{amsmath,amssymb,amsfonts}
\usepackage{graphicx}
\usepackage{textcomp}
\usepackage{xcolor}
\usepackage{gensymb} 
\newtheorem{definition}{Definition} 
\usepackage{algorithm}
\usepackage{algorithmic}
\usepackage[multiple]{footmisc}
\let\labelindent\relax

\usepackage[shortlabels]{enumitem}

\usepackage{comment} 
\usepackage{float} 
\usepackage{longtable} 
\usepackage{enumitem} 

\begin{document}
%
\title{Fine-grained Conflict Detection of IoT Services}

\author{\IEEEauthorblockN{Dipankar Chaki}
\IEEEauthorblockA{School of Computer Science\\
The University of Sydney\\
Sydney, Australia\\
dipankar.chaki@sydney.edu.au}
\and
\IEEEauthorblockN{Athman Bouguettaya}
\IEEEauthorblockA{School of Computer Science\\
The University of Sydney\\
Sydney, Australia\\
athman.bouguettaya@sydney.edu.au}
}

\maketitle

\IEEEoverridecommandlockouts
\IEEEpubid{\begin{minipage}{\textwidth}\ \\[12pt] \centering
  \copyright 20XX IEEE.  Personal use of this material is permitted. Permission from IEEE must be obtained for all other uses, in any current or future media, including reprinting/republishing this material for advertising or promotional purposes, creating new collective works, for resale or redistribution to servers or lists, or reuse of any copyrighted component of this work in other works.
\end{minipage}}

\begin{abstract}
We propose a novel framework to detect conflicts among IoT services in a multi-resident smart home. A fine-grained conflict model is proposed considering the functional and non-functional properties of IoT services. The proposed conflict model is designed using the concept of entropy and information gain from information theory. We use a novel algorithm based on temporal proximity to detect conflicts. Experimental results on real-world datasets show the efficiency of the proposed approach.
\end{abstract}


\IEEEpeerreviewmaketitle

\section{Introduction}
Internet of Things (IoT) is the networked interconnection of everyday objects that are equipped with ubiquitous intelligence \cite{xia2012internet}. IoT is opening tremendous opportunities for some cutting-edge applications due to the rapid advancement of technologies such as Wireless Sensor Networks (WSNs), Radio Frequency Identification (RFID) tags, mobile phones, and actuators \cite{xia2012internet}. \textit{Smart home} is one of the preeminent application domains for IoT. A smart home is equipped with IoT devices that monitor the use of various ``things'' in the home to provide \textit{efficiency} and \textit{convenience}\cite{huang2018discovering}. 

The concept of IoT is very much aligned with the \textit{service paradigm} \cite{bouguettaya2017service}. Each ``thing'' is expected to have a set of purposes (a.k.a. functionalities) delivered with a quality of service (a.k.a. non-functional attributes). We leverage the service paradigm as a framework to define and model the \textit{functional} and \textit{non-functional} properties of smart home devices. In this context, each IoT device is represented as a single \emph{IoT service} \cite{huang2016service}. For example, a TV set in a smart home would be represented as a TV service. The \textit{functional} property of the TV service would be to telecast programs. Examples of \textit{non-functional} properties would include volume, resolution, and connectivity.

We identify two types of smart homes: i) \textit{Single resident} and ii) \textit{Multi-resident}. The distinction between these two types is vital since service conflicts and service recommendations would differ for each type. The focus of this paper is on multi-resident smart homes. Different residents may have different habits in this context, thus having different service requirements, which may lead to \textit{IoT service conflicts} \cite{chaki2020conflict}. Detecting and resolving conflicts in this context is of prime importance. Indeed, a pre-requisite to providing convenience that suits all occupants is to first detect and then resolve conflicts. For example, assume that a resident has a habit of turning the light ``on'' in the living room while watching their favorite program on TV. Assume that the other resident has the habit of turning the light ``off'' while watching the same program on TV. Hence, a service conflict occurs as the light service cannot satisfy the requirements of multiple residents based on their habits.

Conflict detection is the prerequisite of conflict resolution. \textit{Detection of conflicts} is the main focus of this paper. There exist a few works that focus on designing a comfortable smart home without considering IoT service conflicts \cite{fattah2018restful,lin2007multi}. Existing literature on conflict detection is considerably scarce in a context-aware ambient intelligent environment. There are some systems such as CARISMA and Gaia, which consider single and multi-user conflicts regarding a single application \cite{capra2003carisma,ranganathan2003infrastructure}. These systems do not consider multi-user conflicts in a multi-application environment. Some researches have focused on dealing with multi-user preferences for media applications in a smart home \cite{park2005dynamic,shin2009service}. A multitude of applications and services coexist in a smart home such as media applications, light service, AC service.

We identify a few shortcomings of the existing frameworks. Existing approaches do not provide the essential conflict classification scheme for conflict detection. There is a need for a fine-grained conflict classification scheme that can categorize several types of conflicts based on residents' service usage habits. This classification scheme is useful to make decision for conflict resolution. A conflict can be detected a-priori (i.e., likelihood of occurring a conflict) based on residents' service usage patterns. A smart home may adjust appliances' settings to suit the residents' habits if it can detect the conflict before its occurrence. Even if the conflict cannot be resolved automatically, the system may acknowledge its inability to deal with the conflict and may offer alternate services. To the best of our knowledge, existing approaches do not focus on designing an a-priori conflict detection framework for IoT services.

We propose a novel approach for a-priori conflict detection based on residents' service usage habits. Conflict modeling for a-priori conflict detection is a challenging task because different residents have different habits of using IoT services. Different habits denote different functional and non-functional requirements. At first, we build a service usage habit model with a consistency score for each service usage. We use this consistency score or habit score to measure dissimilarity between the residents' usage requirements. We use the concept of \textit{Information Entropy} and \textit{Information Gain} from \textit{Information Theory} to measure the dissimilarity. Information entropy measures the disorder degree of all the instances based on the consistency score. Information gain measures how sparse the consistency scores from one another. We classify three types of conflicts based on entropy and gain value. Then, we develop a conflict detection algorithm using temporal proximity strategy to prune insignificant and loosely correlated service usage requirements. The contributions are summarized as follows:

\begin{itemize}[leftmargin=*]
    \item A fine-grained conflict model that classifies different types of conflicts using entropy and gain score. The scores are calculated based on different residents' different functional and non-functional requirements of IoT services.
    \item A novel a-priori conflict detection algorithm that provides the foundation to resolve conflicts. The algorithm employs a temporal proximity strategy.
\end{itemize}


\section{Motivation Scenario}
We discuss the following two scenarios to illustrate the notion of IoT service conflicts in a multi-resident home.

\begin{figure}[htbp]
\center
\includegraphics[width=\columnwidth]{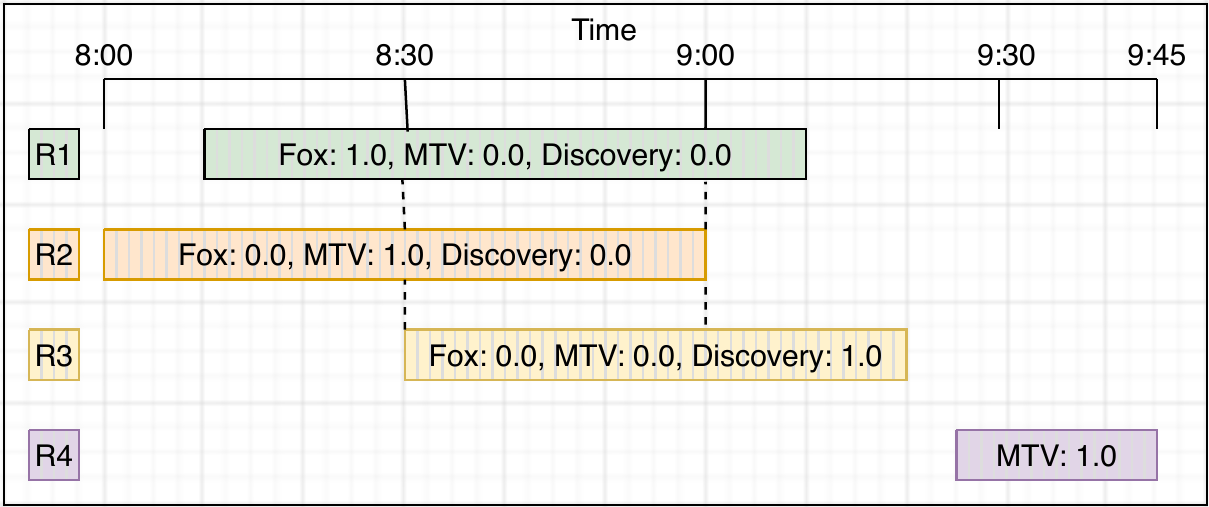}
\caption{Example of a definite conflict.}
\label{fig1}
\end{figure}

\begin{figure}[htbp]
\center
\includegraphics[width=\columnwidth]{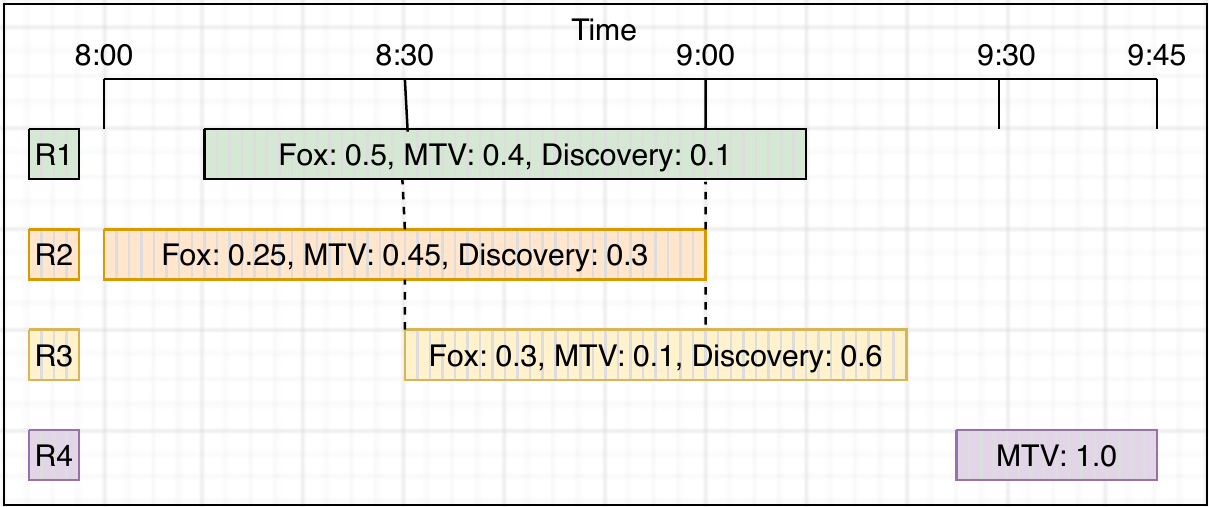}
\caption{Example of a probable conflict.}
\label{fig2}
\end{figure}

\textit{Scenario 1:} Suppose four residents (R1, R2, R3, R4) have four different habits of watching TV. R1 always watches Fox (100\% times); R2 always watches MTV (100\% times); R3 always watches Discovery (100\% times); R4 always watches MTV (100\% times). Here, channel is the functional attribute of the TV service. From Fig.\ref{fig1}, it can be visible that R1, R2, and R3 have overlapping habit intervals in terms of time. If these residents live together, a service conflict will arise because of their completely dissimilar TV watching habits. A TV service cannot telecast more than one channel at the same time duration and location to fulfill different residents’ requirements (assuming the TV doesn’t have multiscreen). If the TV service recommends Fox to the residents, then, R2 \& R3 will be dissatisfied. Since R4 doesn't have an overlapping habit with others, they will not be part of the conflicting situation. From Fig.\ref{fig2}, it can be visible that, R1 watches Fox 50\% times, MTV 40\% times and Discovery 10\% times; R2 watches Fox 25\% times, MTV 45\% times and Discovery 30\% times; R3 watches Fox 30\% times, MTV 10\% times and Discovery 60\% times. R4 always watches MTV (100\% times). Since R1, R2, and R3 have overlapping habit intervals, there is a possibility that a service conflict may arise because of their dissimilar TV watching habits. Detecting this type of conflict beforehand is important, as the TV may recommend alternate services if possible. Even if the TV cannot resolve the conflict automatically, it may acknowledge its inability (i.e., requires human consensus).

\textit{Scenario 2:} Consider a living room equipped with an AC service. A resident (R1) has a habit of keeping AC temperature between 20\degree C and 22\degree C. Another resident (R2) has a habit of keeping AC temperature between 25\degree C and 27\degree C in the living room. The smart AC may know the preferred temperatures of these two residents and may adjust the AC's temperature accordingly. When R1 and R2 stay together in the living room, a conflict arises since the AC service cannot have two different temperature settings at the same time and location. To solve this type of conflict, a good strategy might be to adjust the temperature to the average of the two preference range. For the simplicity of the conflict notion, we assume that the effect of changing the service state is immediate. For example, an AC is executing its operation between 9:00 pm 10:00 pm with 20\degree C temperature. When another request arrives between 10:01 pm and 11:00 pm with 27\degree C temperature, there is no conflict. In practice, the room temperature is not increased by 7 degrees within 1 minute.

\section{IoT Service Model}
We present some preliminaries about \textit{IoT service}, \textit{IoT service event} to demonstrate the notion of \textit{IoT service conflict}. We extend the definition of IoT service and IoT service event from \cite{huang2016service}. We focus on \textit{shareable} IoT services where conflicts may arise. A \textit{shareable} IoT service serves multiple users at the same time and location. Radio, television, DVD, AC, light, heater, and fan are some examples of shared IoT services. A \textit{non-shareable} IoT service serves only one user at a time. Examples are toaster, microwave oven, electric kettle, and washing machine.

\begin{definition}
An \emph{IoT Service ($S$)}, is represented as a tuple of \big \langle \textit{$S_{id}$, $S_{name}$, $F$, $Q$}\big \rangle: \textit{$S_{id}$} is an unique identifier of the service. \textit{$S_{name}$} is the name of the service. \textit{$F$} is the set of \big \{\textit{$f_1$, $f_2$, $f_3$,.......$f_n$}\big \} where each $f_i$ denotes the function offered by a service. \textit{$Q$} is the set of \big \{\textit{$q_1$, $q_2$, $q_3$,.......$q_m$}\big \} where each $q_j$ denotes a QoS attribute of a service.
\end{definition}

\begin{definition}
An \emph{IoT Service Event ($SE$)}, is an instantiation of a service. When a service manifestation occurs (i.e., turn on, turn off, increase, decrease, open, close), an event records the state of the service. A service event can be represented as a tuple of \big \langle \textit{\{$S_{id}, F, Q\}, T, L, U$}\big \rangle: \textit{$S_{id}$} is the unique identifier of the service that has some functional ($F$) and non-functional ($Q$) properties. \textit{$T$} is the execution time of the service. It is represented as a set  \{\textit{$T_s$,$T_e$}\} where $T_s$ denotes the start time and $T_e$ denotes the end time. \textit{$L$} and \textit{$U$} are the location and user of the service, respectively.
\end{definition}


An example of the IoT service event is \big\langle \textit{\{5, \{telecasting programs\}, \{35 dB, 50 nits $\infty$\}\}, \{07:45, 08:45\}, living room, 3}\big\rangle. Here, 5 is the id of the TV service by which it can be uniquely identifiable. 07:45 is the service start time, and 08:45 is the service end time. The execution time of the service is 08:45-07:45=1 hour. During the execution time, it's functionality is telecasting programs (i.e. channels). \{35 dB, 50 nits, $\infty$\} represents the non-functional properties such as volume (35 dB), brightness (50 nits) and capacity ($\infty$) of the service. Living room denotes the location where the service operates its functionality. 3 is the unique identifier of the resident who is using the service.

An IoT service, $S$, is typically associated with a set of functional and non-functional properties. A service can be used distinctively by different residents. Different residents' service usage requirements can be captured from IoT service events, $SE$. An IoT service event demonstrates how a resident is using a particular service, along with time and location. The history of service events is stored in a database called IoT service event sequences, $SES$. Different residents may have different habits (i.e., different service usage requirements) which may cause a conflict, $Conf$. The objective of the paper is to identify a function $F(S,SES)$, where $Conf \approx F(S,SES)$ that can detect a-priori conflict in granular level using service-related and usage-related data.


\section{IoT Service Conflict Detection Framework}

\begin{figure}[htbp]
\center
\includegraphics[width=\columnwidth]{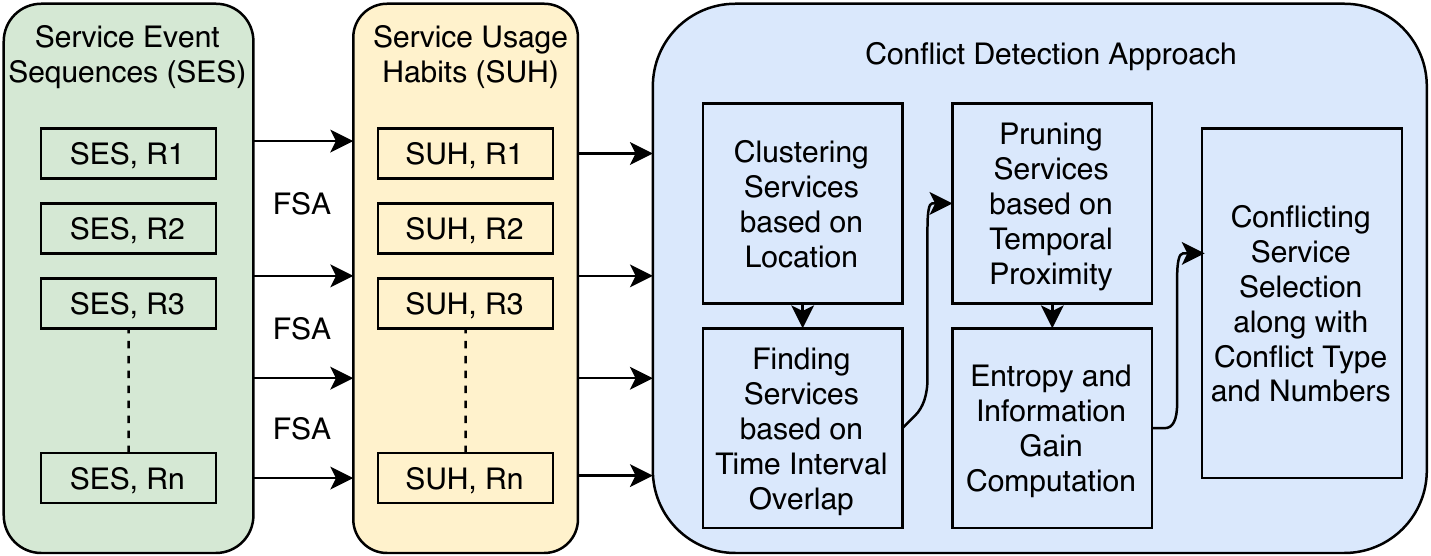}
\caption{IoT service conflict detection framework.} 
\label{framework}
\end{figure}

The overall architecture of the proposed IoT service conflict detection framework, as depicted in Fig. \ref{framework}, is composed of three modules: i) Service Event Sequences, ii) Service Usage Habit, iii) Conflict Detection Approach. Service event sequences represent the service usage history of the residents. Service usage habit model has been built from the previous usage patterns. Then, the usage habit model is fed into the conflict detection module to capture different types of a-priori conflicts.

\subsection{Service Usage Habit Model}
Usually, residents interact with IoT services for various household chores. These interactions are recorded as IoT service event sequences ($SES$). We can get user's service usage habit by analyzing these event sequences. When people use a service frequently and periodically (a.k.a, consistently), that denotes their habits. Based on the consistency score, we define fuzzy service attribute as follows:

\begin{definition}
\emph{Fuzzy Service Attribute ($FSA$)}, is a tuple \big \langle \textit{$FSA_{id}$, $FSA_{name}$, $S_{id}$, $FSA_{val}$}\big \rangle: \textit{$FSA_{id}$} is the unique identifier of the service attribute, \textit{$FSA_{name}$} is the name of the attribute, \textit{$S_{id}$} stands for the service id, \textit{$FSA_{val}$} is a set of pairs representing all possible values of the attribute with their consistency scores.
\end{definition}

For example, a resident always watches sports channel on TV and another resident sometimes watches news channel (60\% time) and sometimes watches geography channel (40\% time). Based on their habits, a fuzzy service attribute for the former is \big\langle \textit{56, channel, 5, \{star sports, 1.0\}}\big\rangle. Here, 56 is the unique identifier of the FSA. Channel is the name of the attribute. 5 is the identifier of a TV service and channel attribute is a part of the TV service. Star sports is the value of the channel attribute. 1.0 denotes that the resident always watches the sports channel (100\% time). For the latter, the fuzzy service attribute becomes \big\langle \textit{56, channel, 5, \{\{Fox, 0.6\}, \{Discovery, 0.4\}\}}\big\rangle. FSA is generated by FSCEP (Fuzzy Semantic Complex Event Processing) which produces complex events with their consistency score \cite{jarraya2016fscep}.

Then, we define the concept of service usage habit of a resident extending the model described in \cite{wang2015habit}. Incorporating fuzzy service attribute with the service usage patterns, we define service usage habit as follows:

\begin{definition}
\emph{Service Usage Habit ($SUH$)}, is a tuple \big \langle \textit{$U$,  $FSA$, \{$SUH_{st}\pm\varepsilon_{st}$, $SUH_{et}\pm\varepsilon_{et}$\}, $L$}\big \rangle: \textit{$U$} is the owner of the habit, \textit{$FSA$} is a single fuzzy service attribute or a set of fuzzy service attributes, \textit{$SUH_{st}\pm\varepsilon_{st}$ and $SUH_{et}\pm\varepsilon_{et}$} stands for the habit start time and end time respectively, \textit{$L$} stands for the location.
\end{definition}

For instance, a service usage habit can be represented as \big\langle \textit{3, \{56, channel, 5, \{\{Fox, 0.6\}, \{Discovery, 0.4\}\}\}, \{19:45$\pm10$, 20:45$\pm20$\}, living room}\big\rangle. Here, 3 is the unique identifier of the owner of the habit. 56 is the unique identifier of the FSA. Channel is the name of the attribute that belongs to a TV service whose id is 5. The resident has a habit of watching Fox 60\% time and Discovery 40\% time between 19:45 and 20:45. $\pm10$ denotes that the resident usually starts watching TV between 19:35 and 19:55. $\pm20$ denotes that the resident usually stops watching TV between 20:25 and 21:05. TV is located in the living room.

A resident may have simple habit and complex habit. When a resident has a habit based on a single service usage, it is regarded as a simple habit. The above example illustrates a simple habit scenario. When a resident has a habit based on more than one service usages, it is regarded as a complex habit. For example, when a resident enjoys movie at the TV, he prefers the light to be off. In this case, he has a habit consists of two services (both TV service and light service). In a smart home, different residents may have different habits for service usage which may cause a conflict. This module is not the core of our contributions; however, it produces the input ($SUH$) for the conflict detection module, which holds the core contributions of the present work.

\subsection{Conflict Classification}
We use the consistency score generated from the service usage habit to estimate the conflict likelihood of the IoT services. \textit{IoT Service Conflict} occurs when a service cannot satisfy the requirements of multiple users at the same time duration and location. Requirements are gathered from service usage habits of different residents. Each habit has a start time and end time along with some functional and non-functional attributes of IoT services. When these habits contain same fuzzy service attribute and have overlapping time periods, they are compared with one another to see the dissimilarity of the service requirements. The more dissimilarity represents more probability of having conflicts. Given two service usage habits ($SUH_i, SUH_j$), the following conditions have to be satisfied to be considered as a conflicting situation.

\begin{itemize}
    \item $(FSA_{SUH_i} \simeq FSA_{SUH_j}) \cap (L_{SUH_i} \simeq L_{SUH_j})$, denoting that two habits have the same service attribute and they are located at the same place.
    \item $(SUH_{i_{st}}, SUH_{i_{et}}) \cap (SUH_{j_{st}}, SUH_{j_{et}})) \neq \emptyset$, meaning that two service usage habits are invoked simultaneously and there is a temporal overlap between them. We use Allen's temporal relation to detect overlapping events \cite{allen1994actions}.
    \item $U_{SUH_i} \neq U_{SUH_j}$, denoting that two habits are invoked by two different users.
\end{itemize}

A habit instance of a resident may have multiple FSAs. Each FSA has a value and its consistency score. For example, a resident may have a habit of watching Fox (channel) 60\% time and Discovery (channel) 40\% time; always (100\% time) keeping volume 30-35 dB while enjoying TV service. This information can be stored in a FSA like $\langle \{\{Fox, 0.6\}, \{Discovery, 0.4\}\}, \{[30-35], 1.0\} \rangle$. Dissimilarity score between different residents' habits is calculated for each attribute. For instance, one dissimilarity score for channel attribute; one dissimilarity score for volume attribute. We measure dissimilarity ($DSim$) using the \textit{information entropy} and \textit{information gain} concepts from the \textit{information theory}. The information entropy measures the disorder degree of all instances based on their consistency score and the information gain measures how sparse those consistency scores from one another. We classify three types of conflicts using the relationship between information entropy and information gain. Equation \ref{entropy} and Equation \ref{gain} correspondingly represent the calculation of information entropy and information gain.

\begin{equation}
Entropy\;E(X) = -\sum p(X)\log p(X)
\label{entropy}
\end{equation}

\begin{equation}
Gain\; G(X,Y)= E(X)-E(X|Y)
\label{gain}
\end{equation}

The following tables can be generated from Fig. \ref{fig1} and Fig. \ref{fig2} considering two cases from scenario 1. For the betterment of the presentation, we only show the overlapping habit instances' (i.e., eliminating R4's habit instance since it is not overlapped with others) consistency score of the channel attribute in the tables.

\begin{table}[htbp]
  \centering
 \caption{\small Consistency Score of Figure 1}
  \label{table1}
    \begin{tabular}{|c|c|c|c|c|}
      \hline
      Resident & Fox  & MTV & Discovery & Total \\
      \hline
      R1 & 100  & 0 & 0 & 100 \\
      \hline
      R2 & 0  & 100 & 0 & 100 \\
      \hline
      R3 & 0  & 0 & 100 & 100 \\
      \hline
      Total & 0 & 0 & 300 & 300 \\
      \hline
    \end{tabular}
 \caption{\small Consistency Score of Figure 2}
  \label{table2}
    \begin{tabular}{|c|c|c|c|c|}
      \hline
      Resident & Fox  & MTV & Discovery & Total \\
      \hline
      R1 & 50  & 40 & 10 & 100 \\
      \hline
      R2 & 25  & 45 & 30 & 100 \\
      \hline
      R3 & 30  & 10 & 60 & 100 \\
      \hline
      Total & 105 & 95 & 100 & 300 \\
      \hline
    \end{tabular}
\end{table}

Each row of the table represents the probability distribution for each resident's habit. For Table \ref{table1}, three residents have completely different habits of watching TV. R1 always watches Fox, R2 always watches MTV and R3 always watches Discovery. we calculate $E(Channel)$ and $G(Channel | User)$ using Equation \ref{entropy} and Equation \ref{gain}. We get the values as follows: $E(Channel)=1.59$ and $G(Channel | User)=1.59$. The maximum gain score can be equal to the entropy score. The maximum entropy score is calculated as: $E_{max} \simeq \log_2 (n)$. Here, $n$ is the number of possible values for each attribute. For instance, channel attribute has three possible values (i.e. Fox, MTV, Discovery). Therefore, the maximum entropy score can be $\log_2(3) \simeq 1.59 $. Since, the gain score is equal to the maximum entropy score, their dissimilarity is very high. This is a case where conflict will happen definitely. When the value of a consistency score is zero, we assume the value is very close to zero, not exactly zero. Otherwise, $\log_2(0)$ would become undefined. For Table \ref{table2}, R1 watches Fox 50\% times, MTV 40\% times and Discovery 10\% times; R2 watches Fox 25\% times, MTV 45\% times and Discovery 30\% times; R3 watches Fox 30\% times, MTV 10\% times and Discovery 60\% times. For this table, entropy score is 1.59 and gain score is 0.20. The gain score is greater than zero which means there is sparsity between the consistency scores. It represents the dissimilarity between their watching habits. Since, there is dissimilarity, there is a probability of having a conflict.

We define three types of conflicts based on the relationship between information entropy and information gain. When residents have high dissimilar habits of IoT service usage, we define it as a \textit{Strong Conflict (SC)}. When the dissimilarity is medium, we call it as a \textit{Tau Conflict (TC)} (i.e. leaning towards conflict). When the dissimilarity is very low (a.k.a almost have similar habits of service usage), we define it as a \textit{Weak Conflict (WC)}. Table \ref{table3} presents several types of conflicts. Here, $G$ is the gain score, $E_{max}$ is the maximum entropy score and $n$ is the possible number of value for each attribute. This fine-grained classification is important, because, conflict resolution module will try to resolve strong conflict first, then, tau conflict and finally, it will deal with weak conflict. Conflict resolution is out of scope of this paper. However, this fine-grained conflict classification is the foundation of the conflict resolution module.
\begin{table}[htbp]
\center
\caption{Conflict Classification}\label{table3}
\begin{tabular}{|l|l|}
\hline
Conflicts &  Condition\\
\hline
Strong Conflict (SC) & $G >= \frac{E_{max}}{2}$ \&\& $G <= E_{max}$ \\
Tau Conflict (TC) & $G >= \frac{E_{max}}{2^n}$ \&\& $G <= \frac{E_{max}}{2}$ \\
Weak Conflict (WC) & $G >= 0$ \&\& $G <= \frac{E_{max}}{2^n}$ \\
\hline
\end{tabular}
\end{table}

\subsection{Conflict Detection Approach}
Our proposed algorithm detects fine-grained service conflicts before it actually happens. The proposed approach has three phases: Phase 1 is the pre-processing stage for the conflict detection. Phase 2 and 3 are illustrated in Algorithm 1 and 2, respectively.
\subsection*{Phase 1: Pre-processing}
Some service event sequence data cannot be used directly. So, the data need to go through some pre-processing steps such as statistical binning and value stabilization. The pre-processing stage aims to convert the events into a form that can be used in the service usage habit module. It considers different attributes of services and convert their values into some categorical states to make it more meaningful. The steps of pre-processing stage are explained as follows:

\paragraph{Statistical Binning} We construct the consistency score of each fuzzy service attribute ($FSA$) (Definition 3) based on categorical values. However, there are some attributes that have numerical values. Therefore, we apply a statistical method called data binning. It takes the continuous numerical values and puts them into multiple categories. We use a dynamic programming approach to get the optimal bin \cite{mishra2020alternate}.

\paragraph{Value Stabilization} For some service attributes, there might be several values that are advertised within a short period of time where only the final value is relevant. For example, browsing through TV channels before settling down at a final channel. In this work, we only consider the final settled down value while measuring service usage habits. At the end of the pre-processing stage, we calculate service usage habits ($SUH$) using the model described in \cite{wang2015habit}. All the $SUH$ of all residents are stored in a database ($DB$) which is the input of Algorithm 1.

\subsection*{Phase 2: Finding Overlapping Service Usage Habits} 
This phase starts with \textit{dividing the search space}. A smart home usually has some locations such as a kitchen, a bedroom, a living room, and a bathroom. Each service is located in a search space. Therefore, each service usage habit is associated with a search space. We cluster service usage habits ($SUH$) that are located in the same location (lines [2-10] in algorithm 1). When a TV service and a DVD service are located in a living room, $SUH$ cluster of the living room is $LSUH_{living} = \big \langle SUH_{TV}, SUH_{DVD} \big \rangle$.

\begin{algorithm}[b!]
\small
\caption{Overlapping Service Usage Habits Selection}\label{alg:algorithm1}
\begin{algorithmic}[1]
\REQUIRE
$DB$ // a database consists of all $SUH$
\ENSURE
$OSUH$ // overlapping service usage habits

\STATE {$LSUH = \emptyset, TM = \emptyset, SV = \emptyset, OSUH = \emptyset$}

\item[] // Clustering services located in a same location
\FOR{\textbf{each} $suh_i$ in $DB$}
\FOR{\textbf{each} $l_j$ in $suh.l$}
\FOR{\textbf{each} $s_k$ in $suh.S$}
\IF{$s_k.l$ is equal to $l_j$}
\STATE $LSUH_i \leftarrow insert(s_k)$
\ENDIF
\ENDFOR
\ENDFOR
\ENDFOR
\item[] // Selecting overlapping service usage habits
\FOR{\textbf{each} $l_i$ in $LSUH$}
\FOR{\textbf{each} $s_j$ in $l_i$}
\STATE $TM \leftarrow addTimeInterval(s_j.T_s, s_j.T_e)$
\ENDFOR
\ENDFOR
\STATE $SV \leftarrow sort(TM)$
\STATE $OSUH \leftarrow overlap(SV)$
\RETURN OV

\end{algorithmic}
\end{algorithm}

For each service located in a location, in terms of time, a set of overlapping habit instances is defined as \textit{Overlapping Service Usage Habit} ($OSUH$). It is represented as $\{\langle SUH_{i_{st}}, SUH_{i_{et}}\rangle,$ $\langle SUH_{j_{st}}, SUH_{j_{et}}\rangle$ ... $\langle SUH_{n_{st}}, SUH_{n_{et}}\rangle\}$. $n$ represents number of habit instances. $SUH$ is defined in Definition 4 and $st, et$ stands for habit start time and end time, respectively. For example, Fig. \ref{fig1} represents three overlapping habit instances of R1, R2 and R3. In the $OSUH$, habit of R4 is not considered as it is not overlapping with other residents' habits. By ordering all elements in $SUH$ in a non-decreasing order based on its associated time information $st, et$, we can transform $SUH$ into the following representation:
\begin{equation*}
    SUH = \langle SUH_{seq}, T\rangle = \Bigg\{   
    \begin{matrix}
    \alpha_1 ... \alpha_i ... \alpha_{2n}\\ 
    t_1 ... t_i ... t_{2n}
    \end{matrix}
    \Bigg\}
\end{equation*}
    
Here, $SUH_{seq} = \alpha_1...\alpha_i...\alpha_{2n}$ is a symbol sequence and $\alpha_i = SUH_j^*$ (* can be + or - based on habit start and end time), $T = \{ t_1,...,t_i,...,t_{2n}$ is the time information. Overlapping Service Usage Habit ($OSUH$) is the output of algorithm 1. $OSUH$ in Fig. \ref{fig1} can be represented as: 
\begin{equation*}
\Bigg\{   
    \begin{matrix}
    SUH_2^+ & SUH_1^+ & SUH_3^+ & SUH_2^- & SUH_1^- & SUH_3^-\\ 
    8:00 & 8:10 & 8:30 & 9:00 & 9:10 & 9:20
    \end{matrix}
    \Bigg\}
\end{equation*}

\subsection*{Phase 3: Detection of Conflicts}
The input of this algorithm is the set of overlapping service usage habits generated from algorithm 1. There are some habit instances that overlap for a small amount of time. For example, one resident has a habit of watching TV between 6:30 pm and 7:00 pm. Another resident has a habit of using the same service between 6:55 pm and 7:30 pm. The overlapping time period is very small (only 5 minutes, i.e., between 6:55 pm and 7:00 pm). We use temporal proximity strategy to filter out some loosely related habit instances.
\paragraph{Temporal Proximity} Temporal proximity technique for evaluating the distance between time-interval data is adopted from \cite{shao2016clustering}. For each service usage habit, $SUH_i = (SUH_{i_{st}}, SUH_{i_{et}})$, we use a function $f_i$ with respect to $t$ to map the temporal aspect of $SUH_i$. Habit start time and end time are represented with $SUH_{i_{st}}$ and $SUH_{i_{et}}$, respectively. $f_i$ is formalized in Equation \ref{proximity}.
\begin{equation}
    f_i(t) = 
    \begin{cases}
      1, & t \in [SUH_{i_{st}}, SUH_{i_{et}}] \\
      0, & otherwise
    \end{cases}
    \label{proximity}
\end{equation}

We generate a set of functions ${f_1, f_2, ... f_n}$ corresponding to the service usage habit instances ($SUH$). The temporal proximity ($temp_{prox}$) for all the overlapping habit instances is calculated by Equation \ref{temporal proximity}.

\begin{equation}
    temp_{prox} = \frac{\int_{t_1}^{t_{2n}}\sum_{i=1}^{n}f_i(t)dt}{(t_{2n}-t_1).n}
    \label{temporal proximity}
\end{equation}


Here, $t_1$ and $t_{2n}$ are the first and the last time information of overlapped habits from $OSUH$ and $n$ is the number of habit instances. Let's assume two residents (R1, R2) have two habits of watching TV. R1 watches TV between 20:00 and 21:00; R2 watches TV between 20:45 and 21:45. The temporal proximity of these two habits can be calculated as $\frac{(20:45-20:00)+(21:00-20:45).2+(21:45-21:00)}{(21:45-20:00).2} = 0.57$. Let's assume another scenario where a resident (R3) has a habit of watching TV between 18:00 and 19:00. Another resident (R4) has a habit of watching TV between 18:10 and 19:10. The temporal proximity of these two habits can be calculated as $\frac{(18:10-18:00)+(19:00-18:10).2+(19:00-19:10)}{(19:10-18:00).2} = 0.86$. The latter scenario has higher temporal proximity than the former one. Thus the latter case is considered to be more prone to have a conflicting situation.

\begin{algorithm}[t!]
\small
\caption{Detection of Service Conflicts}\label{alg:algorithm2}
\begin{algorithmic}[1]
\REQUIRE
$OSUH, \mu$ // temporal proximity threshold, $\mu$
\ENSURE
$CN$,$CT$,$CS$

\STATE ${CN = 0, CT = \emptyset, CS = \emptyset, E = 0, G = 0}$

\STATE $OSUH \leftarrow prune(\mu)$
\FOR{\textbf{each} $osuh_i$ in $OSUH$} 
\FOR{\textbf{each $fsa_j$ in $osuh_i.FSA$}}
\STATE $E \leftarrow computeEntropy(fsa_j)$
\STATE $G \leftarrow computeGain(fsa_j)$
    \IF{$G >= (max(E)/2) \&\& G <= max(E)$}
        \STATE $CT \leftarrow Strong Conflict$
        \STATE $CN_{CT} += 1$ 
        \STATE $CS_{CT} \leftarrow insert(fsa_j)$
    \ENDIF
    \IF{$G >= (max(E)/(2^n)) \&\& G <= (max(E)/2)$}
        \STATE $CT \leftarrow Tau Conflict$
        \STATE $CN_{CT} += 1$ 
        \STATE $CS_{CT} \leftarrow insert(fsa_j)$
    \ENDIF
    \IF{$G >= 0 \&\& G <= (max(E)/(2^n))$}
        \STATE $CT \leftarrow Weak Conflict$
        \STATE $CN_{CT} += 1$
        \STATE $CS_{CT} \leftarrow insert(fsa_j)$
    \ENDIF

\ENDFOR
\ENDFOR
\STATE return $CN$,$CT$,$CS$
\end{algorithmic}
\end{algorithm}

\section{Experimental Results and Discussion}
\subsection{Experimental Setup}
The proposed approach has been evaluated using a real dataset collected from the Center for Advanced Studies in Adaptive Systems (CASAS) \cite{cook2012casas}. We use java programming language, and the experiment is performed on a 3.20GHz Intel(R) Core(TM) i7-8700 CPU and 16 GB RAM under Windows 10 64-bit Operating System. There are a few multi-resident activity datasets available. However, those datasets are not useful as they do not have any conflicting situations of IoT service usage. Multi-resident activity datasets reflect compromises of tenants interacting with services. Thus not showing conflicts. In contrast, records of single-resident interactions with IoT services show the actual preferences of an individual for what conflicts may arise. With that respect, we use the service interaction records of four individual residents. The dataset contains different types of sensors. They are battery level sensors, magnetic door sensors, light switches, light sensors, infrared motion sensors, and temperature sensors. In this study, we consider each \textit{sensor} as an \textit{IoT service}. Table \ref{table4} shows the description of all the attributes in our dataset.

\begin{table}[t!]
\caption{Description of the dataset attributes.}
\label{table4}
\begin{tabular}{|l|l|p{6.7cm}}
\hline
\textbf{Attributes} &  \textbf{Description} \\
\hline
Date &  The service execution date\\ \hline
Time &  The service execution time\\ \hline
Sensor & \parbox{6.7cm}{Name of the sensors such as door sensors, light switch sensors, motion sensors, light sensors, temperature sensors}\\\hline
Status & \parbox{6.7cm}{ON, when the service starts executing, and OFF, when the service stops executing}\\
\hline
\end{tabular}
\end{table}

The dataset contains two months of usage data between June 15, 2011, and August 14, 2011. The dataset only contains the on/off time of the sensors. We consider ``ON" as a service start time and ``OFF" as a service termination time. Other attributes, such as functional and non-functional properties are absent in this dataset. We augment a dataset to test our approach based on QoS attributes of IoT services.

\subsection{Performance Evaluation}
\begin{figure}[htbp]
\begin{center}
\includegraphics[width=0.85\columnwidth]{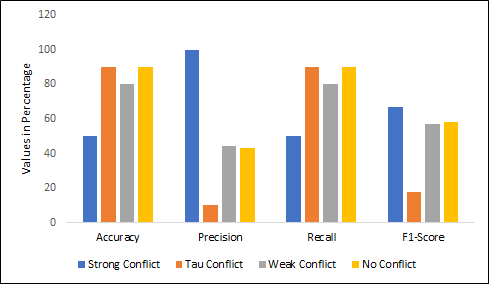}
\caption{Accuracy, precision, recall, f1-score in detecting conflicts.}
\label{fig4}
\end{center}
\end{figure}

At first, we annotate our dataset without employing temporal proximity strategy. Therefore, even if there is a small overlapping habit instances in terms of time, we consider it for entropy and gain computation to detect IoT service conflicts. This is our baseline approach. The augmented dataset contains 5000 rows of service events such as TV, AC, and light services' history. We split the dataset into two sets: i) 80\% is used for training the model and ii) 20\% is used for testing purpose. For this experiment, we only consider the data of TV channels to detect conflict a-priori. There are 230 overlapping service usage habits of TV service that needs attention for conflict detection. According to the baseline approach, there are 200 strong conflicts, 10 tau conflicts, 10 weak conflicts and 10 no conflicts. After having the actual value, we test our proposed approach employing temporal proximity technique (Eq. \ref{temporal proximity}). Initially, we set the proximity threshold ($\mu$) as 0.6. Therefore, any habit instance that has less than 0.6 proximity, is automatically discarded. Our approach detects 100 strong conflicts, 9 tau conflicts, 8 weak conflicts and 9 no conflicts accurately. Precision, recall, f1-score and accuracy are calculated to evaluate the performance of the proposed approach \cite{hasan2017empirical}. From Fig. \ref{fig4}, it is visible that the proposed approach captured tau, weak and no conflicts more accurately than the strong conflict. For strong conflict detection, the accuracy is 50\% and it is understandable that it could not capture 50\% strong conflicts because of pruning habit instances that have more inconsistency among them.

\begin{figure}[htbp]
\begin{center}
\includegraphics[width=0.85\columnwidth]{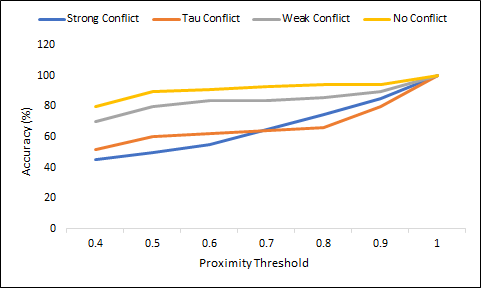}
\caption{Effectiveness of proximity strategy to detect conflicts.}
\label{fig5}
\end{center}
\end{figure}

In the next set of experiment, we adjust the temporal threshold value to observe the accuracy variation in terms of conflict detection. Fig. \ref{fig5} demonstrates that increasing the threshold value can capture different types of conflicts more accurately. It is obvious that when the threshold is small, there are lots of overlapping events that are pruned. When, the events are discarded, the proposed approach cannot detect all the conflicts. However, increasing threshold value denotes that discarding limited number of events, which eventually denotes more capturing ability of conflicts. This is why, when the threshold is 1.0, no events are discarded and all the conflicts are detected (100\%) like baseline approach.

\begin{figure}[htbp]
\begin{center}
\includegraphics[width=0.85\columnwidth]{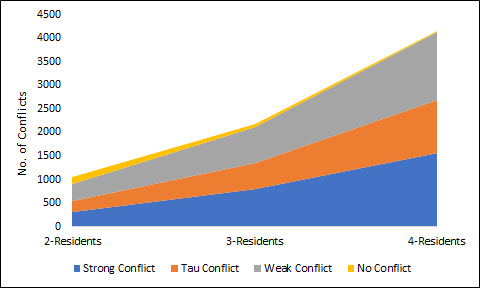}
\caption{No. of conflicts between residents.}
\label{fig6}
\end{center}
\end{figure}

Finally, we conduct another set of experiment to observe how conflicts may increase because of increased usage. We conduct this experiment considering conflicts between 2-residents, 3-residents, and 4-residents, respectively. Fig. \ref{fig6} refers that the number of conflicts increase as the number of residents increase. More residents mean more service usage habits; more service usage mean more service requirements; the more users, the harder to detect and resolve conflicts. 

\subsection{Complexity Analysis}
The proposed approach has a similar polynomial computation complexity like the base line approach. If there are $n$ services and $m$ number of events, the time required to check conflict is $\mathcal{O}(n*m)$ for both approaches. We didn't find any relevant work to compare our approach since the idea of a-priori conflict detection of IoT services for multi-resident smart home is relatively new.

\section{Related Work}
``Conflict is a natural disagreement between different attitudes, beliefs, values, or needs'' \cite{wang2011development}. Relevant research regarding various dimensions of conflict are studied. We found three different dimensions where a conflicts may arise in ambient intelligent environment: (1) source, (2) intervenients, (3) solvability.

A conflict may arise based on four different kinds of \textit{sources}. For example, when multiple users concur over a given resource (e.g., TV), it is known as a \textit{resource-level} conflict \cite{lalanda2017conflict}. When various applications concur over a resource (e.g., building administration applications attempting to control a room's lighting), it is known as an \textit{application-level} conflict \cite{tuttlies2007comity}. When conflict arises because of different contextual policies, it is known as a \textit{policy level} conflict \cite{hu2011semantic}. For example, a user listens music through his smartphone device inside a library, it violates silence policy of the library. When there are diverse user inclinations in a similar setting (e.g., one user wants to peruse with the lights to be at full limit, and another user likes to stare at the TV with the light to be at half limit), a conflict may occur and it is known as \textit{profile level} conflict \cite{carreira2014towards}.

Conflicts may occur as a result of \textit{intervenients} \cite{yagita2015application}. A conflict may occur in a single-resident home because of having conflicting intentions (e.g., comfort vs energy saving). A conflict may emerge between two residents when they use a resource simultaneously. A conflict may happen between occupants and space, where a user's activities strife with any settled space strategies (e.g., a user's cell phone ringing in a room with a quietness approach).
 
Finally, conflict can be categorized by their \textit{solvability} \cite{goynugur2016automatic}. The best case scenario is conflict avoidance. When conflict is detected before its actual occurrence, it can be solved a-priori. Even if the system cannot resolve the conflict, it may acknowledge its inability. None of the above mentioned works detect conflict a-priori. They all detect conflict during its actual occurrence. Additionally, all of them did not consider extracting habits/preferences from previous service usage history. We use previous data to model residents' service usage habits and based on that habits, we develop a conflict detection framework that can capture conflict in advance before its actual occurrence.




\section{Conclusion and Future Work}
We design a fine-grained conflict model that captures different types of IoT service conflicts before its occurrence. The framework provides a foundation for conflict resolution and to recommend convenient and efficient IoT services to the residents. The experimental results show the effectiveness and efficiency of the proposed conflict detection approach. Our future work focuses on the resolution of different types of conflicts and building a recommender system for multi-resident smart homes.

\def\IEEEbibitemsep{3pt plus 1pt}
\bibliographystyle{IEEEtran}
\bibliography{FineGrained}

\end{document}